\documentclass[fleqn,10pt]{wlscirep}

\usepackage[utf8]{inputenc}
\usepackage[T1]{fontenc}
\usepackage[export]{adjustbox}
\usepackage[justification=justified]{caption}

\newcommand{\mbf}[1]{\mathbf{#1}}

\newcommand{\av}[1]{\left\langle #1 \right\rangle}

\begin{document}

\title{Berezinskii-Kosterlitz-Thouless phase induced by dissipating quasisolitons}

\author[1,*]{Krzysztof Gawryluk}
\author[1]{Miros{\l}aw Brewczyk}
\affil[1]{Wydzia{\l} Fizyki, Uniwersytet w Bia{\l}ymstoku, ul. K. Cio{\l}kowskiego 1L, PL-15245 Bia{\l}ystok, Poland}
\affil[*]{k.gawryluk@uwb.edu.pl}

\begin{abstract}
We theoretically study the sound propagation in a two-dimensional weakly interacting uniform Bose gas. Using the classical fields approximation we analyze in detail the properties of density waves generated both in a weak and strong perturbation regimes. While in the former case density excitations can be described in terms of hydrodynamic or collisionless sound, the strong disturbance of the system results in a qualitatively different response. We identify observed structures as quasisolitons and uncover their internal complexity for strong perturbation case. For this regime quasisolitons break into vortex pairs as time progresses, eventually reaching an equilibrium state. We find this state, characterized by only fluctuating in time averaged number of pairs of opposite charge vortices and by appearance of a quasi-long-range order, as the Berezinskii-Kosterlitz-Thouless (BKT) phase. 
\end{abstract}
\maketitle

\section{Introduction}
Sound waves carry information on both thermodynamic and transport properties of a medium they propagate through. In classical hydrodynamics, measuring the speed of sound waves and their attenuation gives an access to characteristics of the medium such as the compressibility and viscosity. In quantum hydrodynamics, with superfluids present, the picture is more complex \cite{PitaevskiiStringari}. For example, liquid helium and weakly interacting Bose gas respond to local perturbation in a qualitatively different ways.

Many experiments exploring the phenomenon of sound propagation in ultracold atomic systems have been already performed. Sound waves were studied in harmonically trapped three-dimensional bosonic gases at very low \cite{Ketterle1997,Hoefer06,Chang08} as well as at higher \cite{Meppelink09} temperatures. Sound velocity was measured at resonance in a mixture of fermionic lithium atoms \cite{Joseph07}. Two sound modes propagating at different speeds, according to predictions of two-fluid hydrodynamics, were observed in a resonant Fermi gas \cite{Sidorenkov13}. The sound diffusion in a unitary three-dimensional Fermi gas was investigated and a universal quantum limit of diffusivity was observed in Ref. \cite{Patel19}. Recently, sound propagation and damping were studied in two-dimensional systems -- a weakly interacting Bose \cite{JBeugnon,Christodoulou20} and strongly interacting Fermi \cite{Bohlen20} gases.

In \cite{JBeugnon}, a gas of rubidium-$87$ atoms is confined inside a quasi-2D rectangular potential with hard walls. A density perturbation is introduced by applying a repulsive potential to the cloud of atoms. This additional potential creates a density dip along one direction which propagates at constant speed when the laser is switched off. During this evolution the density perturbation bounces several times off the walls of the box and its velocity is found to be close to the Bogoliubov sound speed. Several attempts have been already undertaken to reproduce results of this experiment \cite{Ota18,Cappellaro18,Singh20}. They all support experimental observation that below the BKT transition the generated sound waves propagate with velocities close to the speed of second sound, predicted by two-fluid hydrodynamic model. They also predict that sound waves can propagate above the BKT transition, in agreement with experiment and in opposition to what is predicted by two-fluid hydrodynamic model with respect to second sound \cite{Ozawa14,Ota18a}.

\section{Sound waves propagation}
In our numerical simulations we first start with analyzing the response of two-dimensional Bose gas to weak perturbation, as in the experiment of Ref. \cite{JBeugnon}. With the help of Metropolis algorithm \cite{Witkowska10,Grisins14,Karpiuk15,Pietraszewicz15,Gawryluk17,jumpKGMB,Singh20} we create an ensemble of initial states, $\psi(\mbf{r})$, for a two-dimensional Bose gas confined in a box potential with periodic boundary conditions. We work within the grand canonical ensemble, so the temperature $T$ and the chemical potential $\mu$ are control parameters. Members of an ensemble of classical fields are drawn with the probability $\sim \exp((\mu N -E)/k_B T)$, where $N=\sum_{\bf k} |\alpha_{\bf k}|^2$ and $E=(\hbar^2/2 m) \sum_{\bf k} {\bf k}^2 |\alpha_{\bf k}|^2 + (g/2 L^2) \sum_{{\bf k},{\bf j},{\bf l}} \alpha_{\bf k}^*\, \alpha_{\bf j}^*\, \alpha_{\bf l}\, \alpha_{{\bf k}+{\bf j}-{\bf l}}$ are the number of particles and the energy of the state, respectively. Here, $g/L^2$ is the two-body interaction energy with $L$ being the length of a square box potential and $\alpha_{\bf k}$ are the amplitudes of expansion of the classical field in the basis of the box eigenfunctions.  

Next, we disturb the cloud of atoms with the protocol similar to that applied in \cite{JBeugnon}. We create a dip in density in one direction, just by multiplying the initial classical field by an appropriate steplike function. The dip is characterized by two parameters: the width $w$ and the depth $d$. Such disturbance, a kind of density imprinting, is then evolved according to the classical field approximation (CFA) prescription. An equation for a disturbed classical field $\psi(\mbf{r},t)$ is \cite{review}
\begin{equation}
i \hbar\, \partial/\partial t\, \psi(\mbf{r},t) = (-\hbar^2/2m\, \nabla^2 +g |\psi(\mbf{r},t)|^2)\, \psi(\mbf{r},t) \,.  
\end{equation} 
In our case the perturbation is initially located at the center of the box. As a consequence we have two density patterns propagating symmetrically outward as in the experiment of \cite{Ketterle1997}. Integrating density along the direction transverse to the sound propagation we are able to identify traveling waves and find their speeds.

\begin{figure}[thb] 
\includegraphics[height=6.cm, width=8cm,center]{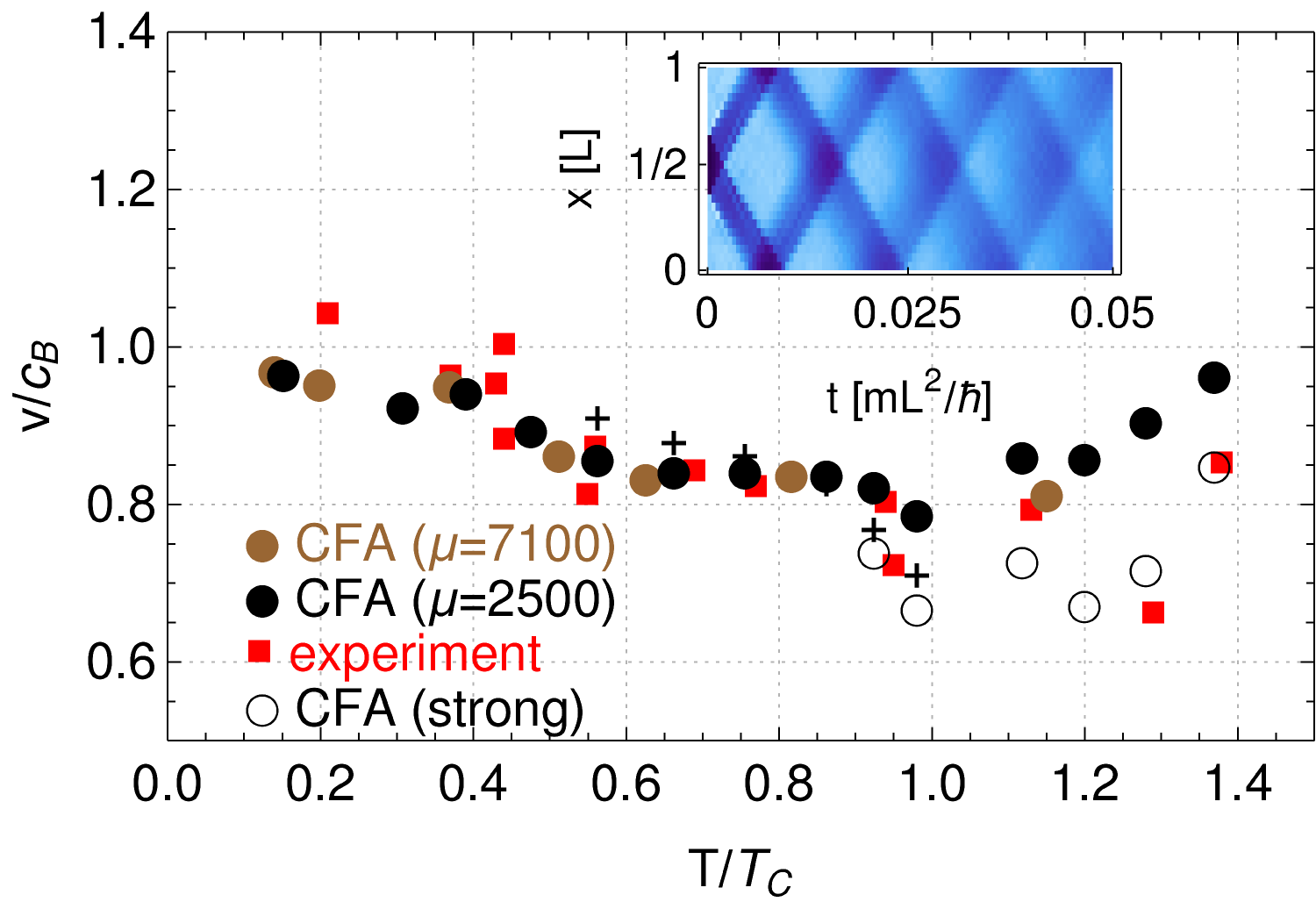} 
\caption{Speed of sound normalized to the Bogoliubov sound speed (at zero temperature) shown as a function of $T/T_c$. Black and brown bullets (obtained at different values of chemical potential, in units of $\hbar^2/m L^2$, as indicated in the legend) come from numerical simulations whereas red squares are experimental results \cite{JBeugnon}. Numerical results are obtained within weak perturbation scheme, with $w/L\approx 0.25$ and $d \lesssim 5/16$. Open circles represent numerical data for stronger disturbance, $w/L \gtrsim 0.25$ and $d > 5/16$. Crosses are the outcome of two-fluid model \cite{PitaevskiiStringari,Ozawa14}, showing the speed of second sound.
Inset: Propagation of sound waves in a box potential. At each time two-dimensional density is integrated along the direction perpendicular to the sound propagation and then averaged over a hundred realizations. The effective one-dimensional density is shown as a function of time. The width of the density depletion equals $w=0.25$ in units of the box length $L$, and the depth of a density dip is $5/16$ of initial density ($N=42275$). The temperature is $T=0.37T_c$. Total evolution time $t=0.05\, mL^2/\hbar$ corresponds to $100\,$ms for $L=38\mu m$. The spatial resolution here is $0.76\mu$m which is $1.58$ of the healing length $\xi=\hbar/\sqrt{m n g}$. All figures were created with the help of Wolfram Mathematica (version 12.1, https://www.wolfram.com/mathematica/).}
\label{Fig1}
\end{figure}

To get the velocities of density waves we decompose the density integrated along the direction perpendicular to the wave propagation, which is $y$ axis in our case, via the Fourier transform at each instant of time:
$n(x,t)=\av{n} + \sum_{j=\pm 1,\pm 2,...} A_j(t) \exp(j\,2\pi x/L)$.
Here, $\av{n}$ is the average density along $x$ axis.
Then we do time analysis of $A_j(t)/A_j(0)$ coefficients -- we fit real (or imaginary) part of them to an exponentially damped sinusoidal function 
$\,e^{-\Gamma_j t/2} \left[ \Gamma_j/(2\omega_j) \sin(\omega_j t) + \cos(\omega_j t)\right]$.
Focusing on lower energy modes we obtain the velocity of density waves as $v=\omega_j/k_j$, where $k_j$ is the wave vector ($k_j=\pm j2\pi/L$, since periodic boundary conditions are applied).

In Fig. \ref{Fig1} we summarize our results, showing velocities of density waves in units of the Bogoliubov speed $c_B=\sqrt{g N/L^2 m}$ and assuming the interaction strength is $g=0.16\, \hbar^2/m$. The temperature is given in units of $T_c=2\pi n \hbar^2/[m k_B \ln{(380\, \hbar^2/m g)}]$, which is the calculated critical temperature for the BKT phase transition \cite{Prokofev01}. Two sets of equilibrium states, corresponding to different values of chemical potential (as indicated in the legend), are used for the analysis. Numerical values of the speed of sound waves (black and brown bullets, black circles) remain in a good agreement with experimental data (red squares \cite{JBeugnon}) in the whole range of studied temperatures. Most of the data in Fig. \ref{Fig1} were obtained assuming a density imprinting parameters: $w/L=0.25$ and $d=5/16$. Some of the points (brown bullets) were calculated in the regime of very weak perturbation, $d=1/16$. Close to the transition and for the normal phase we double results showing also the response of the system to stronger initial disturbance (black circles, for which $d > 5/16$). Numerical data clearly demonstrate that the response of the system depends on the strength of initial density imprinting. Similar observation was reported in Ref. \cite{Chang08}, where the wave fronts propagation through the condensate was studied after it was split by a strong laser beam.

Additionally, we put in Fig. \ref{Fig1} values of the speed of second sound, coming from the two-fluid model \cite{PitaevskiiStringari,Ozawa14,Ota18a}. Within this framework one can calculate a value of the speed of second sound, $c_2$, from the equation
\begin{equation}
c^4-\left( \frac{1}{mn\kappa_S}+\frac{n_s T \tilde{s}^2}{n_n \tilde{c}_V} \right) c^2+
\frac{n_s T \tilde{s}^2}{n_n \tilde{c}_V} \frac{1}{mn\kappa_T}=0 \,,
\label{c2two-fluid}
\end{equation}
where $n_s$, $n_n$, and $n$ $(=n_s+n_n)$ are superfluid, normal, and total densities, respectively.  $\kappa_T$ ($\kappa_S$) is the isothermal (adiabatic) compressibility, $\tilde{s}$ is the entropy, and $\tilde{c}_V$ -- the specific heat at constant volume, both per unit mass. For temperatures such that $k_B T > g\, n$, a good approximation to the velocity of second sound is given by the expression $c_2=\sqrt{n_s/n}/\sqrt{m n \kappa_T}$ \cite{Ozawa14}. Within CFA we can calculate both the superfluid density (from current-current correlations \cite{jumpKGMB}) and the isothermal compressibility. The latter is obtained from the fluctuation-dissipation relation $(\langle N^2 \rangle - \langle N \rangle^2)/ \langle N \rangle^2 = k_B T \kappa_T / V$ \cite{Huang}. The speed of second sound, as a function of $T/T_c$, is plotted in Fig. \ref{Fig1} by black crosses and the values are close to experimental (and CFA) results.

In CFA approach the classical field is expanded in the basis of modes which are macroscopically occupied -- in our case these are simply the plane waves \cite{review}. Dynamical equations for modes amplitudes (expansion coefficients) can be written in the form of a set of nonlinear differential equations. At each instant of time, these amplitudes determine fully the classical field. Hence, we can watch the density (square modulus of the classical field) with arbitrary high spatial resolution. In particular, this resolution can be high enough to resolve the healing length scale.

\begin{figure}[thb] 
\begin{center}
\includegraphics[width=14.0cm,center]{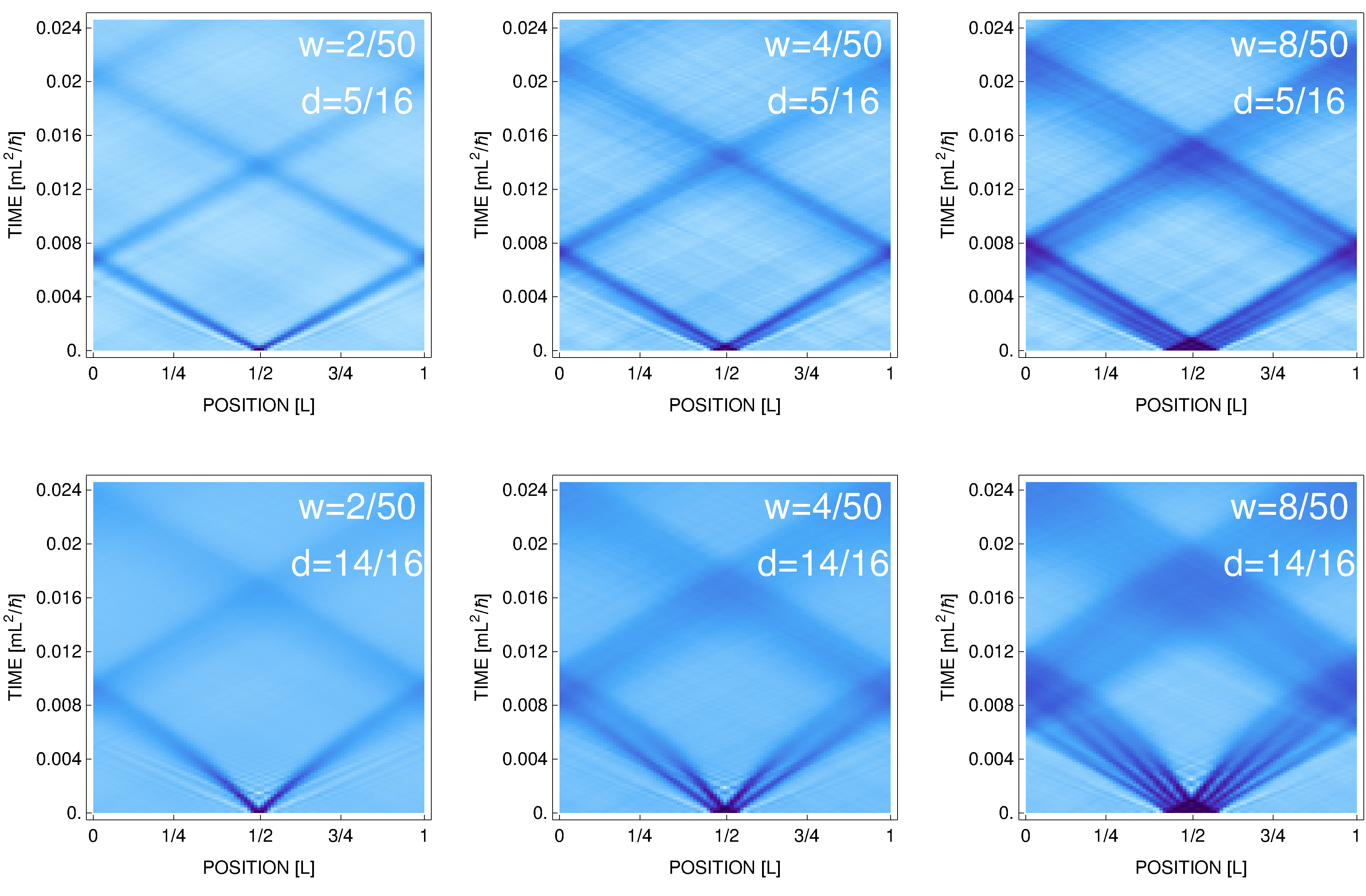} 
\end{center}
\caption{Time evolution of initially perturbed equilibrium state. Two-dimensional density of the system is first integrated along one direction, next averaged over approximately $100$ realizations, and then shown as a function of time. Perturbation parameters, the width $w$ and depth $d$ are given in the frames. Temperature is $T=0.37T_c$. Total evolution time $t=0.025\, mL^2/\hbar$ corresponds to $50\,$ms for $L=38\mu m$. Here, the spatial resolution is $1.64$ of the healing length. }
\label{timeEvo2}
\end{figure}

\section{Quasisolitons}

New observations are possible in a strong perturbation regime with high resolution employed. Interesting things happen when the density imprinting parameters, the width $w$ and the depth $d$ of the initial depletion region, are being increased. First of all, we find that the visible density waves, in fact, consist of several thinner structures, which are characterized by slightly different velocities, see Fig. \ref{timeEvo2}. For weaker initial perturbations they rather form a single beam (Fig. \ref{timeEvo2}, upper row), whereas for stronger ones we clearly see signs of nonlinear behavior -- lines in time-position plots (Fig. \ref{timeEvo2}, lower row) are no longer straight. In both cases, internal structures, traveling with slightly different velocities, separate each other. For stronger initial perturbation the multiple density dips accelerate during the motion. At the same time, larger the width of initial perturbation $w$ bigger the number of created waves.
The number of internal thin waves visible inside broad dips in Fig. \ref{timeEvo2} is well described by the formula $w/(4\xi)$, where $w$ is the width of initial perturbation and $\xi$ is the healing length. Factor $4$ arises from a typical width of these structures, see Fig. \ref{timeEvo2}, which turns out to be about $4 \xi$.

\begin{figure}[thb] 
\begin{center}
\includegraphics[width=12cm]{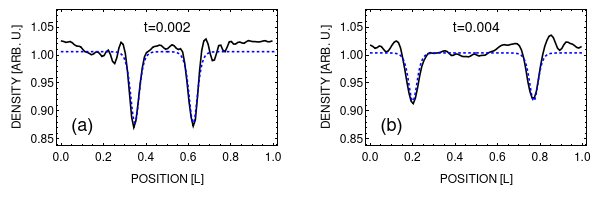}
\end{center} 
\begin{center}
\includegraphics[width=12cm]{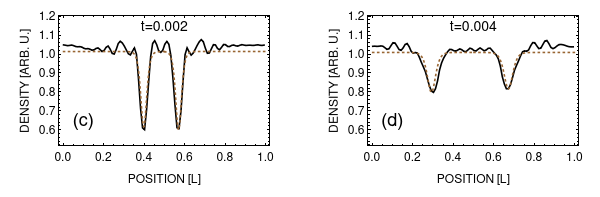}
\end{center}
\caption{Upper row: Density cuts for week initial perturbation ($w/L=2/50, d=5/16$, which corresponds to the most left in upper row frame in Fig. \ref{timeEvo2}) together with two-soliton fits according to dissipative Zakharov-Shabat solution, Eq. (\ref{1Dquasisoliton}), for different times (here, $\dot{q}=0.90\, c_B$). For frames (a) and (b), i.e. before the first collision, $\Gamma=126$ and $\Gamma=169$, respectively. 
Lower row: Density cuts for stronger initial perturbation ($w/L=2/50\,, d=14/16$, which corresponds to the most left in lower row frame in Fig. \ref{timeEvo2}) together with fits according to ``accelarating'' two-soliton solutions of Eq. (\ref{1Dsoliton}), for different times ($\dot{q}=0.77\, c_B$ for (c) and $\dot{q}=0.90\, c_B$ for (d)).}
\label{solitonsFit2}
\end{figure}

Structures presented in Fig. \ref{timeEvo2} survive several crossings and exhibit a moderate change of shapes during the evolution, so one can try to assign some solitonic properties to them. 
Below we exam numerically obtained densities by comparing their profiles with analytical formulas for a single soliton solution of a nonlinear Schr\"odinger equation, known as Zakharov-Shabat solutions with the nonlinear term corresponding to repulsive interactions \cite{Zakharov}
\begin{equation}
|\phi_{ZS}(x,t)|^2 =  
n_0\left[ \frac{\dot{q}^2}{c_B^2}+\left(1-\frac{\dot{q}^2}{c_B^2}\right) \tanh^2\left(\frac{x-q}{\xi}\sqrt{1-\frac{\dot{q}^2}{c_B^2}}\,\right) \right] .  
\label{1Dsoliton}
\end{equation}
Here, $n_0$ is the density of the medium far away from the soliton dip, $q$ and $\dot{q}$ are the position and velocity of the soliton, respectively. The healing length $\xi=\hbar/\sqrt{m n_0 g}$ and $c_B=\sqrt{n_0 g/m}$. The width and the depth of solitonic solution (\ref{1Dsoliton}) is solely determined by soliton velocity, $\dot{q}$. For weak perturbations (top left frame in Fig. \ref{timeEvo2}) density dips propagate with constant velocity but they get shallower in time (due to snake instabilities \cite{Anderson01,Dutton01,Feder00,Tsuchiya08,Ohya19}). Hence, additional damping factor has to be included in Eq. (\ref{1Dsoliton}) and then a dissipative Zakharov-Shabat solution takes the form
\begin{equation}
|\psi(x,t)|^2=\exp(-\Gamma(t)\, t)\, |\phi_{ZS}(x,t)|^2  \,.
\label{1Dquasisoliton}
\end{equation}
On the other hand, for strong perturbations (lower row in Fig. \ref{timeEvo2}) the density waves accelerate in time. In this case, the parameter $\dot{q}$ in Eq. (\ref{1Dsoliton}) should depend on time.

To verify this concept we analyze first the simplest case -- a small initial perturbation in both $w$ and $d$ which leads to single wave moving with constant velocity. In Fig. \ref{solitonsFit2}, frames (a) and (b), we compare numerically obtained density cuts for various times with Eq. (\ref{1Dquasisoliton}), where $\dot{q}$ is fixed by the value of $v$ we get from the time-dependent analysis of $A_1(t)$ coefficient (actually, the numerical results are fitted to the sum of two one-soliton solutions of Eq. (4)). It turns out that both densities match perfectly up to first collision, after which the agreement slowly degrades.
Surprisingly, Eq. (\ref{1Dsoliton}) works well also for strong perturbation, although in this case one has to consider rather an accelerating two-soliton solutions of (\ref{1Dsoliton}), i.e. a solution with time-dependent $\dot{q}$. Frames (c) and (d) prove that an agreement is good, here $w/L=2/50$ and $d=14/16$. Hence, we will be naming observed structures as quasisolitons as opposed to true solitonic objects present in 1D but also in 2D, where they are called the Jones-Roberts (JR) solitons \cite{Jones82,Bongs17}. Similarly to our case (as discussed in the next section), the JR solitons can transform into vortex-antivortes pairs. However, this happens in the limit of vanishing velocity which critically distinguish JR structures from ours. Surprisingly, Eq. (\ref{1Dsoliton}) also works for weak perturbations at longer times, i.e. after first collision of density waves. Density dips become, as time progresses, broader and shallower and their velocity increases towards speed of Bogoliubov sound (similarly to dissipative dynamics of solitons already observed in early experiments on dark solitons created by phase imprinting technique \cite{Burger99} and theoretically discussed in \cite{Fedichev99}).

The excitation protocol we use in numerical simulations is rather a kind of ideal one. It differs from that applied in Ref. \cite{JBeugnon} in two ways. First, we generate the system at equilibrium in a box potential, i.e. without an additional potential repelling atoms out of desired space in the box, and only after that part of the atoms is removed from the sample. Second, the change of the density at the border of perturbed-unperturbed regions is extremely sharp. The second difference, as we checked, does not influence the results in weak perturbation regime. However, when the system is strongly disturbed, multi-quasisoliton structures visible in Fig. \ref{timeEvo2} get less pronounced. The same happens when the system is disturbed and evolved after it is equilibrated in the presence of repelling barrier.

%%%%%%%%%%%%%%%%%%%%%%%%%%%%%%%%%%%%%%%%%%%%%%%%%%%%%%%%%%%%%%%%%%%%%%%%

\section{Emergence of the BKT phase}
\begin{figure*}[h!tb]
\centering
\includegraphics[width=13.0cm]{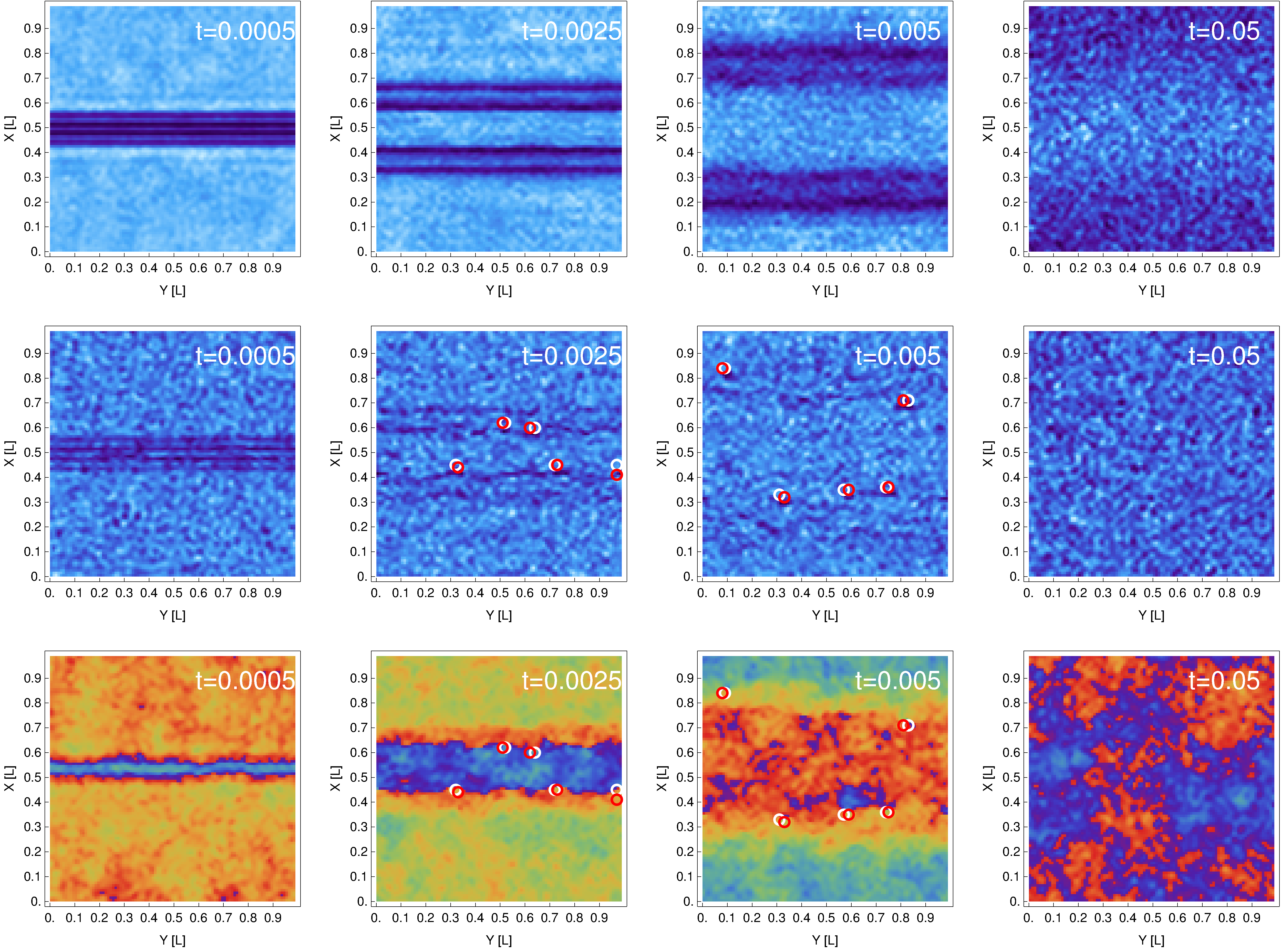}  \\
\includegraphics[width=13.0cm]{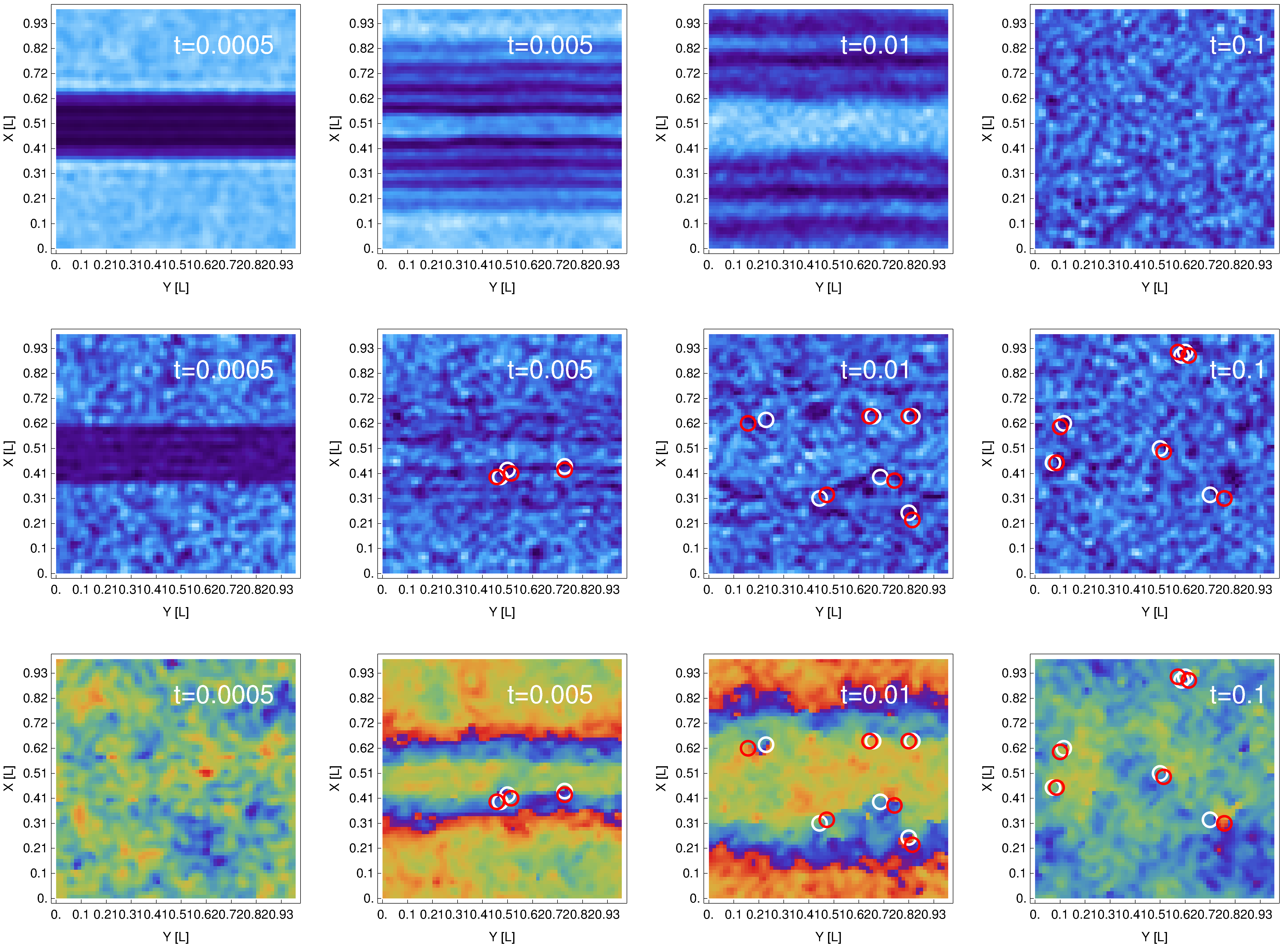}
\caption{First (from the top) row: 2D time-snapshots of a density (averaged over $100$ realizations) for different times as labeled in frames. Second (third) row: 2D densities (phases) for a single realization, with marked white and red circles surrounding elementary vortices with opposite signs. Vortices are identified by calculating the change of the phase around each plaquette of our numerical grid. Here, $w/L=4/50$, $d=14/16$, and $T=0.37T_c$, which corresponds to the middle lower row frame in Fig. \ref{timeEvo2}. Next three rows show the same quantities but for $w/L=12/34$, $d=14/16$, and $T=0.56T_c$ (see Fig. \ref{rhoSdiag}). The spatial resolution is $1.12\mu$m which is $1.39$ of the healing length. }  
\label{density2Dav}
\end{figure*}

\begin{figure}[thb] 
\includegraphics[width=5.5cm]{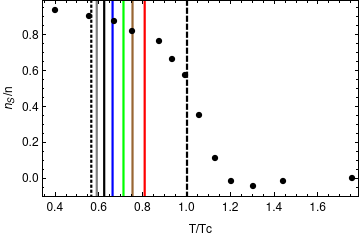} 
\includegraphics[width=5.5cm]{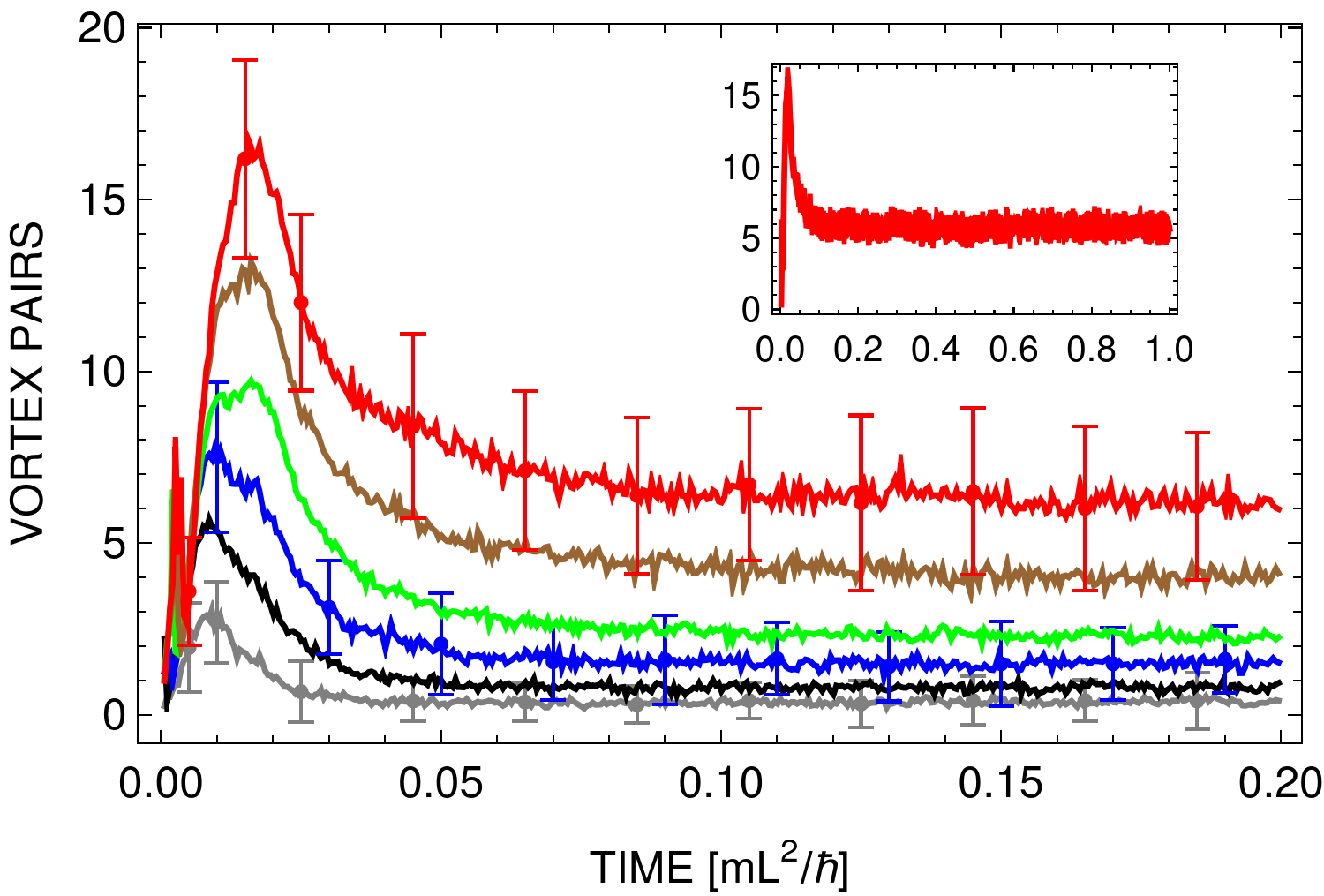} 
\includegraphics[width=5.5cm]{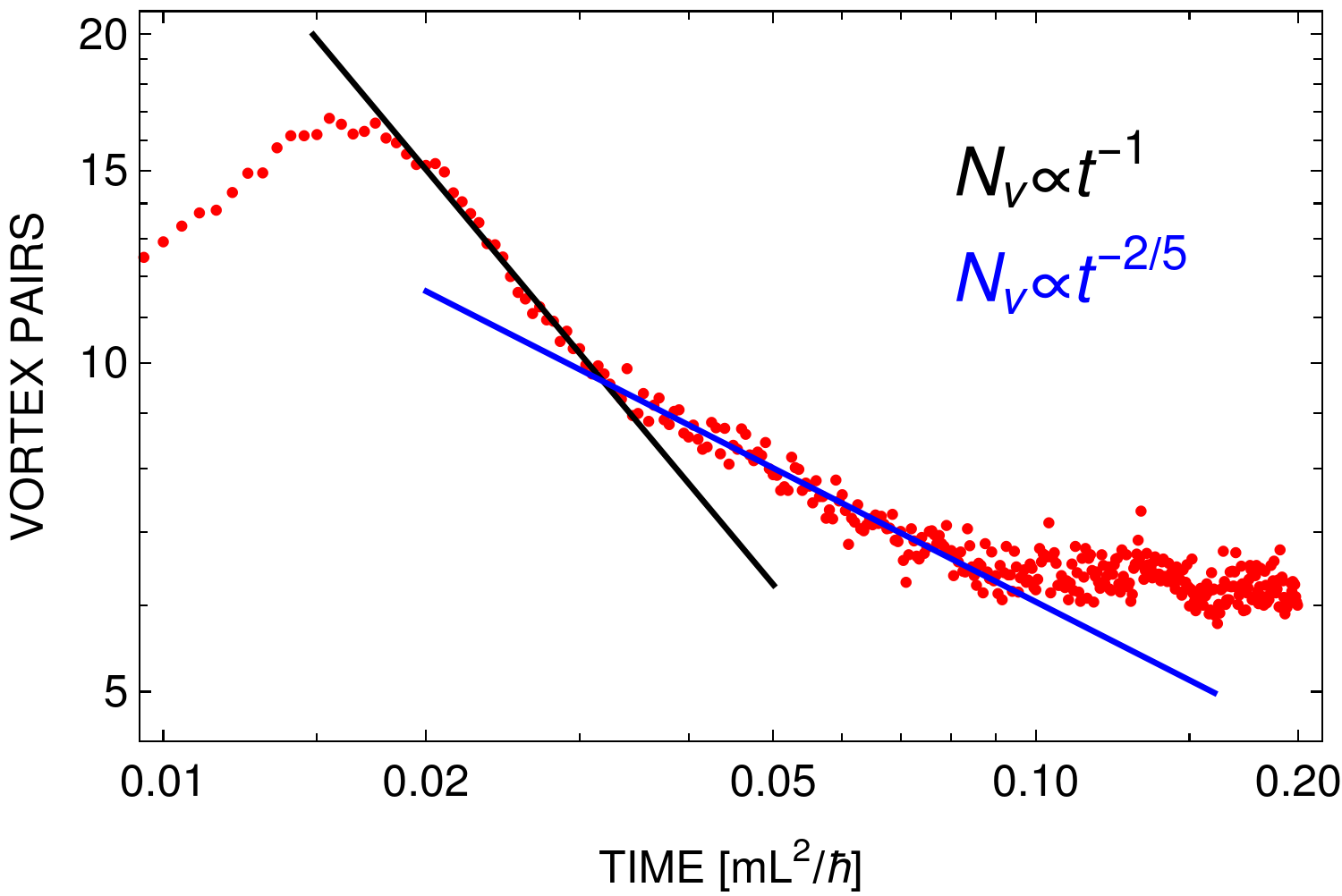}
\caption{Left frame: Superfluid fraction as a function of temperature (black dots) for $\mu=2500\, \hbar^2/m L^2$ (from \cite{jumpKGMB}). Initially, i.e. before perturbation, the system is prepared at temperature $T/T_c=0.56$ (vertical dotted line, most left). After density imprinting with the depth $d=14/16$ and the widths $w/L=(2,4,6,8,10,12)/34$ the system's relative temperature is shifted up (shown by successive vertical lines). The vertical dashed line, most right, shows the transition temperature. Middle frame: Number of pairs of opposite charge vortices (averaged over $100$ realizations; error bars denote the standard deviations) versus time, for various perturbations as in the left frame. Stronger perturbation leads to larger number of vortices at equilibrium, i.e. at longer times and to larger standard deviations. Total evolution time $t=0.20\, mL^2/\hbar$ ($1.0\, mL^2/\hbar$ in inset) corresponds to $400\,$ms ($2\,$s) for $L=38\mu m$. Right frame: Decay of the number of vortex pairs $N_{\rm v}$ as a function of time, here for $w/L=12/34$, exhibiting two characteristic regimes: $t^{-1}$ (black solid line) and $t^{-2/5}$ (blue solid line) dependence. } 
\label{rhoSdiag}
\end{figure}

As observed in simulations, quasisolitons dissipate and eventually breake into elementary vortex pairs (i.e.  pairs of vortices with winding numbers equal to $\pm 1$). It is illustrated in Fig. \ref{density2Dav}, second and fifth rows, where densities for a single realization are shown at different times. Except most left frames almost no trace of density waves is visible, although, while averaged over a hundred realizations, density dips are still detectable for short times (first left three columns in first and fourth rows). Additionally, positions of pairs of opposite charge vortices are marked by color (white and red) circles for a single realization cases. We identify vortices by calculating the phase winding around each lattice plaquette as the system evolves. The number of vortex pairs changes in time but finally, for times longer than $0.1\, mL^2/\hbar$, an equilibrium is established -- see plateau in Fig. \ref{rhoSdiag} (middle frame), exhibiting only fluctuating, after averaging over a hundred realizations, number of vortex pairs. This feature sustains over hundreds of milliseconds as shown in inset in Fig. \ref{rhoSdiag}. The final number of vortex pairs depends on the strength of initial density perturbation, the stronger disturbance the larger number of pairs. For longer times no trace of quasisolitons' positions is visible (see most right column in Fig. \ref{density2Dav}) and motion of pairs of vortices resembles rather that of an ideal gas particles.   

Production of vortex-antivortex pairs in two-dimensional condensates, due to snake instability of initially imprinted dark soliton, was thoroughly discussed in \cite{Verma17}. As demonstrated theoretically in \cite{Singh17}, creation of vortex-antivortex pairs is expected to occur in a laser-stirred two-dimensional Bose gas, as in the experiment of \cite{Desbuquois12}. There, an additional energy is pumped into the system due to stirring the gas at high enough velocity. This energy dissipates via mechanism of creation of vortex-antivortex pairs. Vortex dynamics itself features fast (related to annihilation of vortices of opposite charges) and slow (related to drifting vortices out to the thermal region of the cloud) decay in number of vortices. Remarkably, the first channel is quickly closed and the system exhibits very slow relaxation, on time scales of seconds. The somehow opposite route, which is the crossover from the BKT phase \cite{Kosterlitz73,Kosterlitz74} containing pairs of vortices to a vortex free Bose-Einstein condensate (BEC) in weakly interacting quasi-two-dimensional Bose gas was experimentally realized in \cite{Choi13}.

\begin{figure}[thb] 
\includegraphics[width=7.5cm,center]{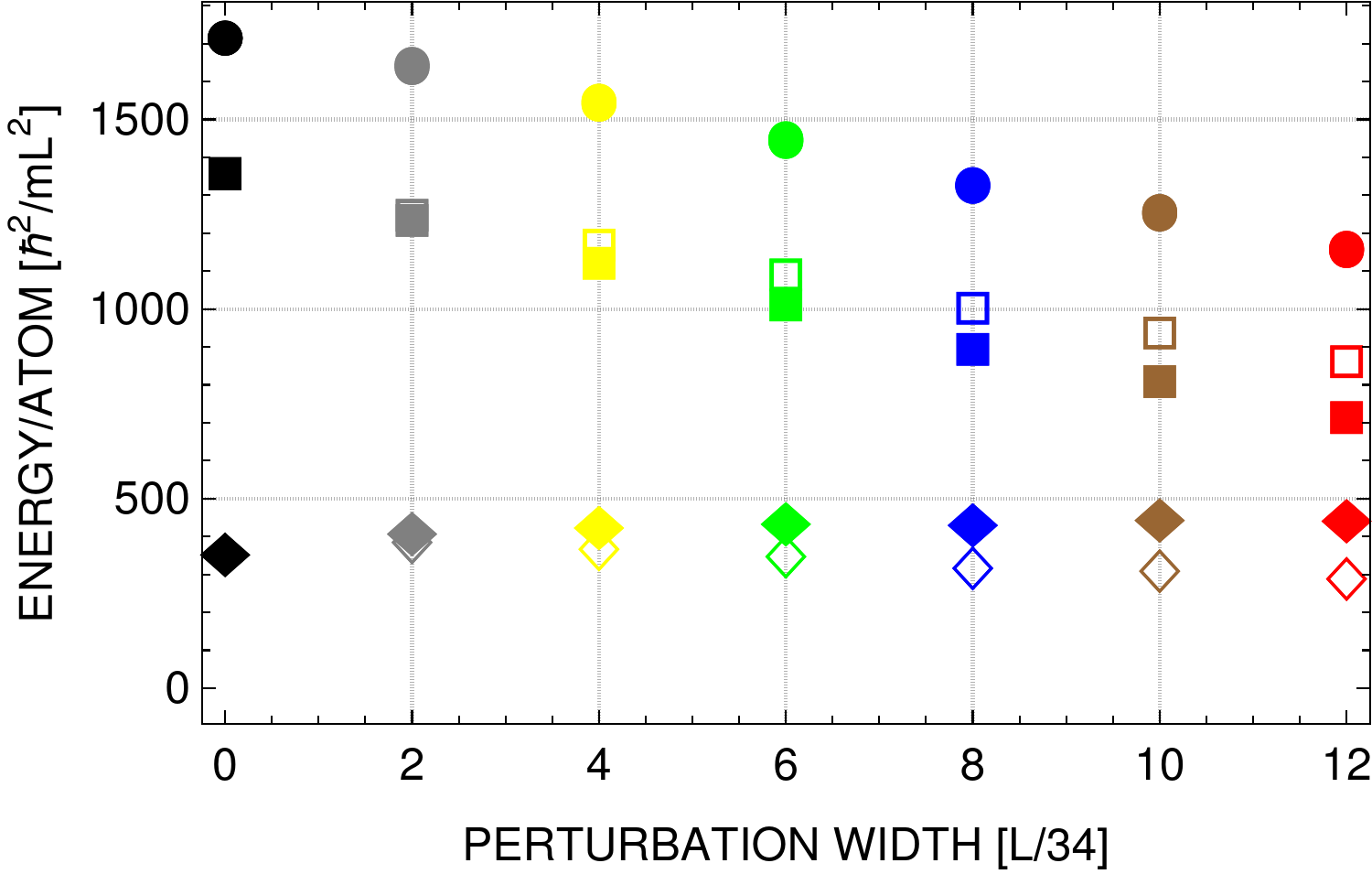} 
\caption{Total (circles), interaction (squares), and kinetic (diamonds) energies per atom as a function of perturbation width, as in Fig. \ref{rhoSdiag}. Open symbols represent initial energies while the solid ones are energies calculated at $t=0.2\,mL^2/\hbar$, deeply in plateau region (total energy is conserved as the system is isolated). A small rise in kinetic energy per atom, which is a measure of temperature of the system, is observed. Then, the increase of temperature $T$ and the reduction of the critical temperature $T_c$ results in increase of the relative temperature $T/T_c$, shown by vertical lines in Fig. \ref{rhoSdiag} (left frame). } 
\label{energies}
\end{figure}

In numerical simulations, initially prepared sample of the Bose gas at equilibrium is modified by immediate removing some part of the atomic cloud. Then, in our case the total energy of the system is decreased and preserved during further evolution as the system is isolated. How the energy per atom is changed by such a perturbation is shown in Fig. \ref{energies}. What is most important is the behavior of kinetic energy per atom, since it determines the temperature of the system. Fig. \ref{energies} tells us that this quantity does not change much. Since the atomic density gets lower, the transition temperature is reduced (see \cite{Prokofev01}). Then effectively, a two-dimensional Bose gas is shifted towards the transition point because $T/T_c$ gets larger.
This supposition is confirmed also by analyzing the properties of the system in plateau region in Fig. \ref{rhoSdiag}, middle frame. It turns out that there exist such values of the chemical potential and temperature that the grand canonical ensemble formalism predicts the average energy and particles number just as those exhibited by the system in plateau region in Fig. \ref{rhoSdiag}.

As visible in Fig. \ref{rhoSdiag}, the number of vortex pairs increases with the strength of perturbation. Left frame in Fig. \ref{rhoSdiag} shows effective ratio $T/T_c$ after disturbance. In the background a superfluid fraction is plotted as a function of relative temperature $T/T_c$ (black dots) for the Bose gas of the chemical potential $\mu=2500\, \hbar^2/m L^2$ (see \cite{jumpKGMB}). The most right line in Fig. \ref{rhoSdiag} (left frame) indicates the transition temperature as argued in \cite{jumpKGMB}, based on consideration of current-current correlations. The superfluid density at the transition temperature represents the superfluid density jump characteristic for the BKT theory as shown by Nelson and Kosterlitz \cite{Nelson77}. What we observe in our simulations is then an opposite route to that realized in \cite{Choi13}, it is rather the crossover from a vortex free BEC to the BKT phase containing pairs of vortices.

\begin{figure*}[htb]
%\begin{figure}[thb] 
\hspace{0.3cm}\includegraphics[width=5.0cm]{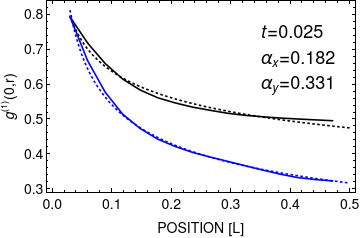} 
\includegraphics[width=5.0cm]{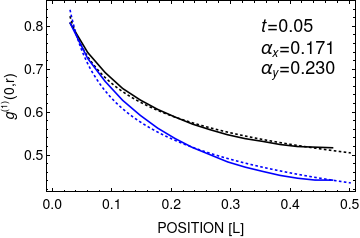} 
\includegraphics[width=5.0cm]{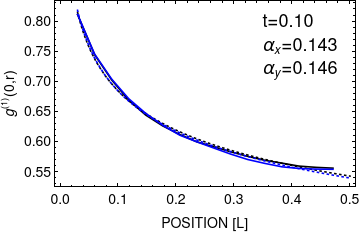} \\
\includegraphics[width=15.70cm]{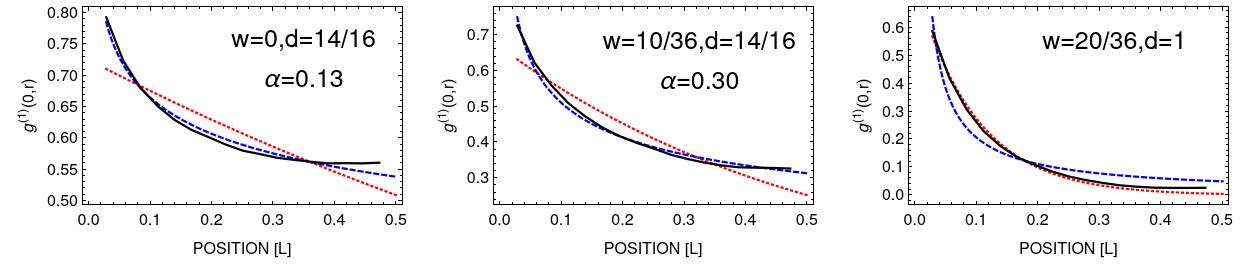}
\caption{Upper row: First-order correlation function, $g^{(1)}(r)$, at various times, showing how the system enters the BKT phase. Initially, the system is prepared at temperature $T/T_c=0.56$. Then, a strong perturbation characterized by $w/L=12/34$ and $d=14/16$ is applied. For times shorter than $0.1\, mL^2/\hbar$ the correlations develop but are not yet spatially uniform (left and middle frames, solid black (blue) lines show correlations along $x$ ($y$) axis). Dashed lines are fits $\sim 1/r^\alpha$ with exponents given in the legends. For longer times (right frame) the system already exhibits quasi-long-range order. Lower row: $g^{(1)}(r)$ function at longer times for $T/T_c=0.8$ and no perturbation (left frame), $w/L=10/36$ and $d=14/16$ (middle frame), and $w/L=20/36$ and $d=1$ (right frame). For stronger disturbance transition to the normal phase with exponential decay of correlations is observed (red dotted curve shows the best exponential fit). }  
\label{g1corfun}
%\end{figure}
\end{figure*}

Presence of vortices in plateau region, Fig. \ref{rhoSdiag} (middle frame), could be a signature of appearance of the BKT phase \cite{Hadzibabic06,Chomaz15}. Here, we actually observe an appearance and persistence of pairs of vortices which is a strong signature for the BKT phase \cite{Kosterlitz73,Kosterlitz74}. As we checked, the average number of vortex pairs does not change over a time of hundreds of milliseconds (see inset in Fig. \ref{rhoSdiag}), suggesting that the system remains frozen in the BKT state. To clarify the situation we calculate the first-order correlation function $g^{(1)}(\mbf{r},\mbf{r'},t)$, defined as a normalized average $g^{(1)}(\mbf{r},\mbf{r'},t)=\av{\psi^{\star}(\mbf{r},t)\, \psi(\mbf{r'},t)} /\av{\psi^{\star}(\mbf{r},t)} \av{\psi(\mbf{r'},t)}$ over the grand canonical ensemble, while the system evolves. According to Kosterlitz and Thouless \cite{Kosterlitz73,Kosterlitz74} it should decay algebraically with a distance in a uniformn two-dimensional Bose gas. We plot the results in Fig. \ref{g1corfun} for a strong perturbation with $w/L=12/34$ and $d=14/16$. We find that already when the number of vortex pairs approaches plateau (see Fig. \ref{rhoSdiag}), the first-order correlation function decays algebraically (Fig. \ref{g1corfun}, left and middle frames in top row). However, the correlations are not yet spatially uniform. They behave as $\sim 1/r^\alpha$ with a distance but with different exponents in $x$ and $y$ directions. However, for $t \gtrsim 0.10\, mL^2/\hbar$ the system starts exhibiting the quasi-long-range order, i.e. $g^{(1)}(\mbf{r},0)$ drops algebraically with a distance, here with exponent equal to $\approx 0.13$ for $t > 0.20\, mL^2/\hbar$. Hence, the system enters the BKT phase. To check the consistency of our simulations we calculate the superfluid density of the system which, according to the BKT theory \cite{Nelson77}, is related to the decay exponent via $\alpha(T)=m k_B T/ 2\pi \hbar^2 n_s(T)$. It gives $n_s(T)/n = (T/T_c)/[\alpha(T) \ln{(380\, \hbar^2/m g)}] \approx 0.8$ and this result agrees with the superfluid fraction of the BKT phase at the temperature $T/T_c \approx 0.8$ (Fig. \ref{rhoSdiag}, left frame, vertical red line), which is the estimated temperature after equilibration in the case of perturbation characterized by $w/L=12/34$ and $d=14/16$. Observation of signatures of vortex-antivortex pairing in, initially perturbed, two-dimensional superfluid Bose gas was already reported in \cite{Seo17}.

Details of the relaxation dynamics of our system, i.e. those exhibited in a period before the plateau region is reached, are depicted in Fig. \ref{rhoSdiag}. We find the number of vortices is decreasing in time. Several experiments, studying two-dimensional turbulent flow in bosonic superfluids, have already reported a decay of vortices after an energy was pumped to the system through transient stirring \cite{Kwon14,Seo17,Gauthier19,Johnstone19}. In \cite{Johnstone19} two distinct regimes have been identified -- the one showing $t^{-2/5}$ and the other exhibiting $t^{-1}$ decay in time of number of vortices. Such vortex annihilation behavior can be understood in terms of collisions involving few vortices. In Ref. \cite{Karl17} it was predicted that vortices annihilate mainly through two- and three-body loss events. These processes result in specific time decay of the number of vortices, governed by $\propto t^{-1}$ and $\propto t^{-2/5}$ dependence, respectively. Experimental results of \cite{Johnstone19} show reasonable agreement with this prediction for some kinds of disturbance protocols. 

Fig. \ref{rhoSdiag} (right frame) clearly shows the existence of two different regimes while the system approaches an equilibrium. We uncover faster ($\propto t^{-1}$) and slower ($\propto t^{-2/5}$) decrease of number of vortices as well. In our case, a slow decay is succeeding the fast one. This must be related to the perturbation protocol we use. It leads to creation of pairs of vortices (vortex dipoles) due to snake instability of quasisolitons rather than to spatially random distributed vortices. Evidently, in the first stage two-body head-to-tail collisions of vortex dipoles are dominant, whereas later on all other relative orientations of vortex dipoles get possible which, effectively, corresponds to the three-body collision events.

Yet another transition can be realized with the perturbation protocol we apply. Now, we start with the two-dimensional Bose gas at higher temperature, $T/T_c=0.8$. The system remains in the BKT phase \cite{jumpKGMB}, see Fig. \ref{g1corfun}, most left frame in bottom row. The first-order correlation function decays algebraically. Then we strongly disturb the Bose gas with $w/L=10/36$ (middle frame) and $w/L=20/36$ (right frame). After stronger perturbation the correlation function starts to decrease exponentially with a distance, the system looses the  quasi-long-range order. Hence, we observe the BKT to normal phase transition in this case.

Moreover, we checked that also in the case of less ideal excitation protocols the long-term evolution exhibits plateau in the number of vortex pairs as in Fig. \ref{rhoSdiag}. Such behavior is expected because the transition temperature $T_c$ is again reduced since the average density is lower than the unperturbed one, due to the presence of additional atoms-repelling barrier.

\section{Summary}
In summary, we have studied the propagation of density waves in a two-dimensional weakly interacting uniform Bose gas. In a weak perturbation regime, the velocities of sound waves agree with the experimentally measured. When the system is strongly disturbed, its response becomes complex. We then identify patterns of multiple density dips, propagating with slightly different velocities. Size of these structures is of submicron and their shape coincides with dissipative Zakharov-Shabat profile. These structures dissipate, changing into pairs of opposite charge vortices. On a scale of hundreds of milliseconds an equilibrium is reached, characterized by only fluctuating in time averaged number of pairs of opposite signs vortices and by appearance of a quasi-long-range order -- the system enters the BKT phase. The BKT phase appears as a result of an increase of system's relative temperature $T/T_c$.

\section*{Acknowledgements}
We are grateful to P. Deuar, M. Gajda, and K. Rz\c{a}\.zewski for helpful discussions. Part of the results were obtained using computers at the Computer Center of University of Bialystok.

\section*{Author contributions statement}
All authors made essential contributions to the work, discussed results, and contributed to the writing of the manuscript. The numerical simulations were performed by K.G.

\end{document}